\documentclass[%
 reprint,
superscriptaddress,
 amsmath,amssymb,
 aps,
prc,
 floatfix,
]{revtex4-2}

\usepackage{graphicx}
\usepackage{dcolumn}
\usepackage{bm}

\begin{document}

\preprint{APS/123-QED}

\title{Evolution of the isoscalar giant monopole resonance in the Ca isotope chain }

\author{
S.~D.~Olorunfunmi,$^{1}$ 
R.~Neveling,$^{2}$
J.~Carter,$^{1}$                    
P.~von Neumann-Cosel,$^{3}$
I.~T.~Usman,$^{1}$
P.~Adsley,$^{1,2,4,5}$
A.~Bahini,$^{1}$
L.~P.~L.~Baloyi,$^{1}$
J.~W.~Br\"{u}mmer,$^{4}$
L.~M.~Donaldson,$^{2}$
H.~Jivan,$^{1}$
N.~Y.~Kheswa,$^1$
K.~C.~W.~Li,$^{4}$
D.~J.~Mar$\acute{\rm i}$n-L$\acute{\rm a}$mbarri,$^{6}$
P.~T.~Molema,$^{1}$
C.~S.~ Moodley,$^{1}$
G.~G.~O'Neill,$^{6}$
P.~Papka,$^{2,4}$
L.~Pellegri,$^{1,2}$
V.~Pesudo,$^{6}$
E.~Sideras-Haddad,$^{1}$
F.~D.~Smit,$^{2}$
G.~F.~Steyn,$^{2}$
A.~A.~Aava,$^{1,2}$
F.~Diel,$^7$
F.~Dunkel,$^7$
P.~Jones,$^{2}$
and
V.~Karayonchev$^7$
\\
$^{1}$\textit{School of Physics, University of the Witwatersrand, Johannesburg 2050, South Africa}\\
$^{2}$\textit{iThemba Laboratory for Accelerator Based Sciences, Somerset West 7129, South Africa}\\
$^{3}$\textit{Institute f\"{u}r Kernphysik, Technische Universit\"{a}t Darmstadt, D-64289 Darmstadt, Germany}\\
$^{4}$\textit{Department of Physics, University of Stellenbosch, Matieland 7602, South Africa }\\
$^{5}$\textit{Institut de Physique Nucl\'{e}aire d'Orsay, IN2P3-CNRS, Universit\'{e} Paris Sud, Orsay, France}\\
$^{6}$\textit{Department of Physics, University of the Western Cape, Bellville 7535, South Africa}\\
$^{7}$\textit{Institute f\"{u}r Kernphysik, Universit\"{a}t zu K\"{o}ln, 50937 K\"{o}ln, Germany}\\
}
\thanks{E-mail: sundayolorunfunmi@gmail.com  (S. D. Olorunfunmi), neveling@tlabs.ac.za (R. Neveling)}

\date{\today}
             
\begin{abstract}
\begin{description}
\item[Background]
Two recent studies of the evolution of the isoscalar giant monopole resonance (ISGMR) within the calcium isotope chain report conflicting results. 
One study suggests that the monopole resonance energy, and thus the incompressibility of the nucleus 
$K_{A}$ 
increase with mass, which implies that $K_{\tau}$, the asymmetry term in the nuclear incompressibility, has a positive value. The other study reports a weak decreasing trend of the energy moments, resulting in a generally accepted negative value for $K_{\tau}$. Differences in the observed trends have been attributed to the use of different techniques to account for instrumental background and the physical continuum.
\item[Purpose] 
An independent measurement of the central region of the ISGMR in the Ca isotope chain is provided to gain a better understanding of the origin of possible systematic trends.
\item[Methods]
Inelastically scattered $\alpha$ particles from a range of calcium targets ($\mathrm{^{40,42,44,48}Ca}$), observed at small scattering angles including 0$^\circ$, were momentum analyzed in the K600 magnetic spectrometer at iThemba LABS, South Africa.  
Monopole strengths spanning an excitation-energy range between 9.5 and 25.5 MeV were obtained using the difference-of-spectra (DoS) technique, adjusted to  allow corrections for the variation of the angular shape of the sum of the $L>0$ multipoles as a function of excitation energy, and compared with previous results that employed multipole-decomposition analysis (MDA) techniques.
\item[Results]
The structure of the $E0$ strength distributions of $^{40,42,44}$Ca agrees well with the results from the previous measurement that supports a weak decreasing trend of the energy moments, while no two datasets agree in the case of $^{48}$Ca.
Despite the variation in the structural character of $E0$ strength distribution from different studies, we find for all datasets that the moment ratios, calculated from the ISGMR strength in the excitation-energy range that defines the main resonance region, display at best only a weak systematic sensitivity to a mass increase. 
\item[Conclusion]
Different trends observed in the nuclear incompressibility are caused by contributions to the $E0$ strength outside of the main resonance region, and in particular for high excitation energies.
While procedures exist to identify and subtract instrumental background, more work is required to characterize and subtract continuum background contributions at high excitation energies.
\end{description}
\end{abstract}

\maketitle

\section{\label{sec:level1}Introduction}

The isoscalar giant monopole resonance (ISGMR) is a
compression mode that can be used to obtain information
about the incompressibility of nuclear matter \cite{Gar18}.   
It was first identified in the late 
1970s \cite{Har77,You77}
and has been extensively studied throughout the 1980s and early 1990s 
at the Texas A\&M University Cyclotron Institute (TAMU) and KVI Groningen \cite{Shl93}.
These experiments employed 
inelastic $\alpha$ and $^{3}$He scattering at very forward angles 
at beam energies between 96 MeV and 130 MeV.
For the past 20 years, the main advances in experimental knowledge
of the ISGMR in stable nuclei originated almost exclusively  from TAMU 
and the Research Center for Nuclear Physics (RCNP)
through small angle (including 0$^{\circ}$) inelastic $\alpha$ scattering 
measurements at 240 MeV and 386 MeV, respectively \cite{Gar18}.
In two mass regions results from systematic studies performed at these two facilities led to opposing conclusions regarding the evolution of the  incompressibility of the nucleus ($K_A$) with mass number.

In the mass region near $A\sim$ 90 Youngblood {\it et al.} \cite{You13,You15,Kri15} 
reported an unexpected variation in the shape of the ISGMR strengths in 
$^{90,92}$Zr and $^{92}$Mo
from experiments carried out at TAMU.
Specifically, the ISGMR energy for $^{90}$Zr was reported to be
1.22 MeV and 2.80 MeV lower than that for $^{92}$Zr and $^{92}$Mo, respectively, resulting
in a value for $K_A$ that increases with mass number.
This anomalous effect was attributed to nuclear shell structure effects,
which is contrary to the generally held notion that the ISGMR and nuclear 
incompressibility are collective phenomena, i.e.\ there is no 
expectation of strong variations in these phenomena due to the internal structure of nuclei \cite{Gar18}.
In response to these interesting findings, Gupta {\it et al.} \cite{Gup16,Gup16b} studied the ISGMR
in the same nuclei at RCNP. 
Nearly identical spectra were observed for the three nuclei, and similar ISGMR strengths were extracted 
leading to the conclusion that the ISGMR energies and strengths for all three
nuclei agree within experimental uncertainty.

The second example of different systematic trends seen from studies at the different facilities
concerns the Ca isotope chain. Results from ISGMR studies at TAMU 
for  $^{40,44,48}$Ca \cite{You01,Lui11,But17} 
showed an increase of the ISGMR centroid energy with mass number \cite{But17}.
This was again attributed to possible nuclear shell structure effects,
and is contrary to expectation according to the hydrodynamic model which predicts 
ISGMR energies decreasing with mass number. 
Such an increase of the value of $K_A$ with mass number would imply that the asymmetry term in the nuclear incompressibility ($K_{\tau}$)
has a positive value.
In an attempt to verify 
such an unusual phenomenon, 
Howard {\it et al.}~\cite{HOWARD2020135185} used the experimental facilities at RCNP
to study the evolution of the ISGMR strength in $^{40,42,44,48}$Ca.
As in the case of the TAMU experiments the MDA technique was employed to extract the ISGMR strength.
In contrast to the TAMU results \cite{You01, Lui11, But17}  
it was found  that the ISGMR strength distributions for the studied Ca isotopes 
follow the generally expected trend of a decrease of the ISGMR centroid energy with mass number increase. 
This would imply that there are no local structure effects, and that a positive value
of $K_{\tau}$ is ruled out.


Gupta {\it et al.} \cite{Gup16} speculated that in the case of $^{90,92}$Zr and $^{92}$Mo 
the reason for the different trends in ISGMR strengths extracted from the different experiments 
lies in the way that the instrumental background and physical continuum
is accounted for by the TAMU and RCNP groups in their respective experiments. 
A similar assertion was later made for the observed discrepancies in the 
case of the Ca isotopes \cite{HOWARD2020135185}.

In light of the potential impact on our understanding of nuclear incompressibility
it is important to establish 
the true nature of the systematic trend, or at the very least to
understand where the observed differences originate from.
Towards this goal we embarked on an independent measurement of the ISGMR strength 
in the calcium isotopes $^{40,42,44,48}$Ca as measured through inelastic $\alpha$ scattering
at the iThemba Laboratory for Accelerator Based Sciences (iThemba LABS) in South Africa.

\section{\label{sec:ExperimentalMethod}Experimental Techniques}

\begin{figure}[ht]
\centering
\includegraphics[width=\columnwidth]{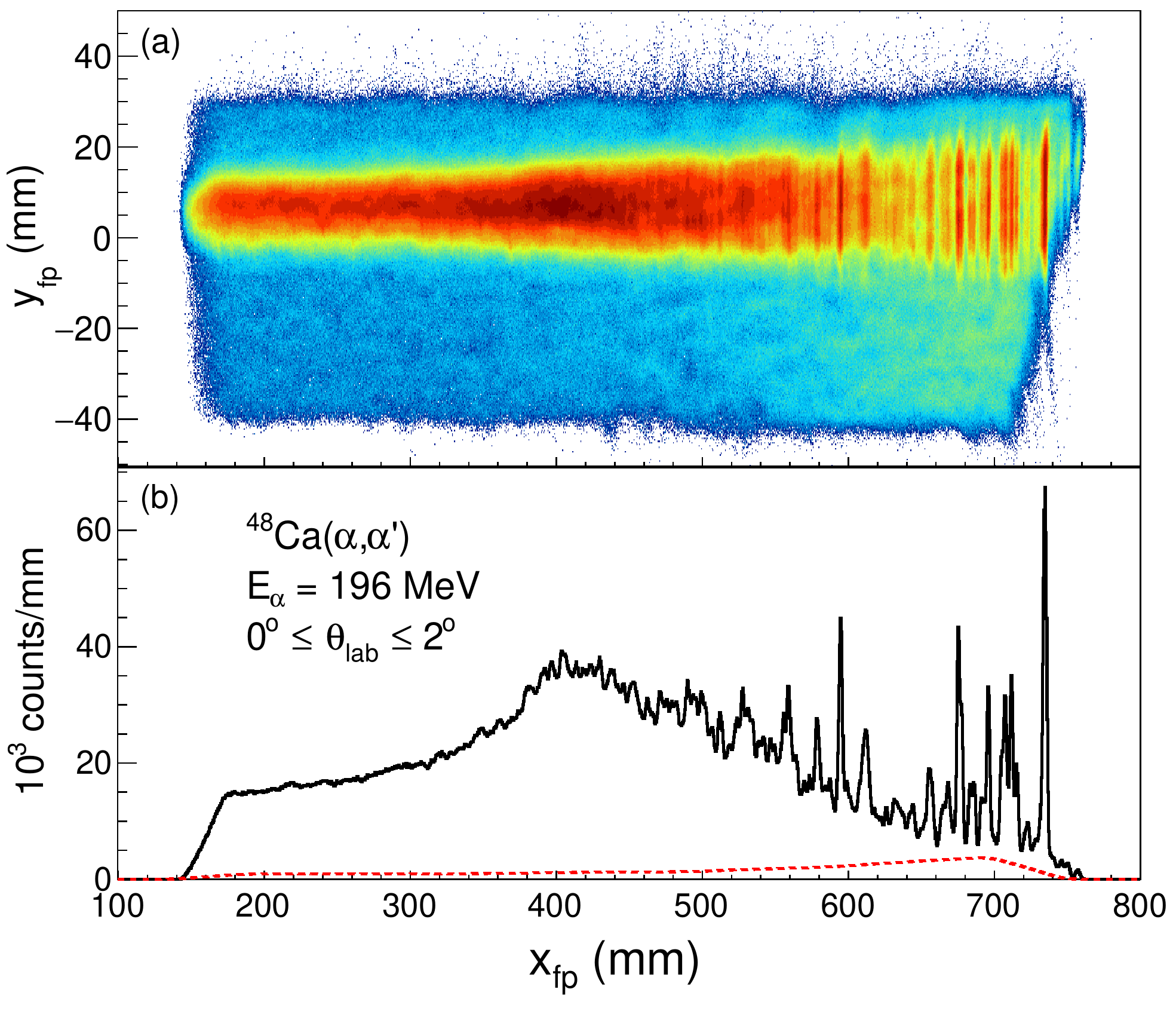}
\caption{
(a) The ($x\textsubscript{fp}$ ,$y\textsubscript{fp}$) distribution of PID selected events 
observed in the focal plane for the  $^{48}$Ca$(\alpha,\alpha^\prime)$ reaction 
at $E_{\alpha}=196$ MeV. A display in color threshold of $>2$ events is imposed.
In panel (b) the (solid) black line represents the 
focal-plane position spectrum for the vertical region from -12 mm to 20 mm.
The (dashed) red line represents the smoothed background obtained when 
selecting $y\textsubscript{fp}$  from -30 mm to -12 mm. 
} 
\label{fig:background}
\end{figure}


Experiments were carried out at the Separated Sector Cyclotron (SSC) facility of iThemba LABS.
A beam of 196 MeV dispersion-matched $\alpha$ particles
was inelastically scattered off a $^{nat}$Ca target as well as $^{42,44,48}$Ca targets with areal densities of 
1.06(5), 1.42(7), 1.54(8) and 1.20(6) mg/cm$^2$, respectively. 
The $^{42,44,48}$Ca targets 
were 98.68\%, 93.71\% and 90.04\% isotopically enriched, respectively.
Scattered particles were momentum analyzed in the high energy resolution K600 Q2D magnetic spectrometer. The horizontal and vertical positions of the scattered $\alpha$ particles at the focal plane ($x\textsubscript{fp}$  and $y\textsubscript{fp}$) were determined using a detector system consisting of two multiwire drift chambers (MWDCs), and a plastic scintillator detector downstream from the MWDCs was utilized to construct the data acquisition trigger. %
Particle identification (PID) was achieved through the standard method of combining information on 
energy deposition in the scintillator with a relative measurement of 
time-of-flight through the spectrometer \cite{Nev11}.

For each target, data were acquired at two spectrometer angles.
In the first measurement the spectrometer
was operated in the 0$^{\circ}$ mode \cite{Nev11}. In this mode the spectrometer is positioned at $\theta$\textsubscript{lab} = 0$^{\circ}$ and both the primary beam and the 
scattered particles enter the spectrometer, 
which has a circular entrance aperture that allows for coverage of $\theta$\textsubscript{lab} $\leq$ 2$^{\circ}$.
Due to the small difference in magnetic rigidity between the
projectiles and scattered particles the beam exits the
spectrometer very close to the MWDCs and the scintillator located in the high-dispersion focal plane of the spectrometer, and is stopped inside a well-shielded Faraday cup four meters downstream from the focal-plane detectors.
In the second measurement the spectrometer was positioned at $\theta$\textsubscript{lab} = 4$^{\circ}$,
covering the scattering angle range $\theta$\textsubscript{lab} = 2 - 6$^{\circ}$. 
In this case the beam was stopped in a Faraday cup adjacent to the  entrance aperture of the spectrometer. In this mode the MWDCs and scintillator were placed at the medium-dispersion focal plane.

Inelastic scattering measurements at very forward scattering angles suffer inevitably from background due to beam halo  and small-angle elastic scattering off the target foil followed by re-scattering 
inside the spectrometer vacuum chamber. Different time-of-flight characteristics associated with beam halo
make it possible to remove such events from the analysis through relevant time cuts.
The remaining unavoidable instrumental background can be quantified and subtracted 
because the reaction products of interest are vertically focused into a narrow band in the focal plane,
while the instrumental background is evenly distributed in the vertical direction.
Such vertically off-focus regions can therefore be used to construct background spectra
as shown in Fig.~\ref{fig:background},
which are subtracted from the focused region.
This method is similar to the procedure followed in experiments performed at RCNP \cite{Gup16,HOWARD2020135185}.

For the 0$^{\circ}$ measurement the spectrometer was operated in a vertically focused ion-optical mode,
resulting in the loss of information regarding the vertical component of the scattering angle. However, in the 4$^{\circ}$
measurements the weak under-focus mode \cite{tam09} was employed whereby the ion optics of the 
spectrometer were adjusted to spread out the inelastically scattered
$\alpha$ particles in the vertical direction, while still allowing for a sizable off-focus
region to be used for background characterization.
In this mode the vertical component of the scattering angle can be determined, which allows for 
the accurate determination 
of scattering angles within the 2$^{\circ}$ - 6$^{\circ}$ angular acceptance.

\begin{figure}[t]
\includegraphics[width=\columnwidth]{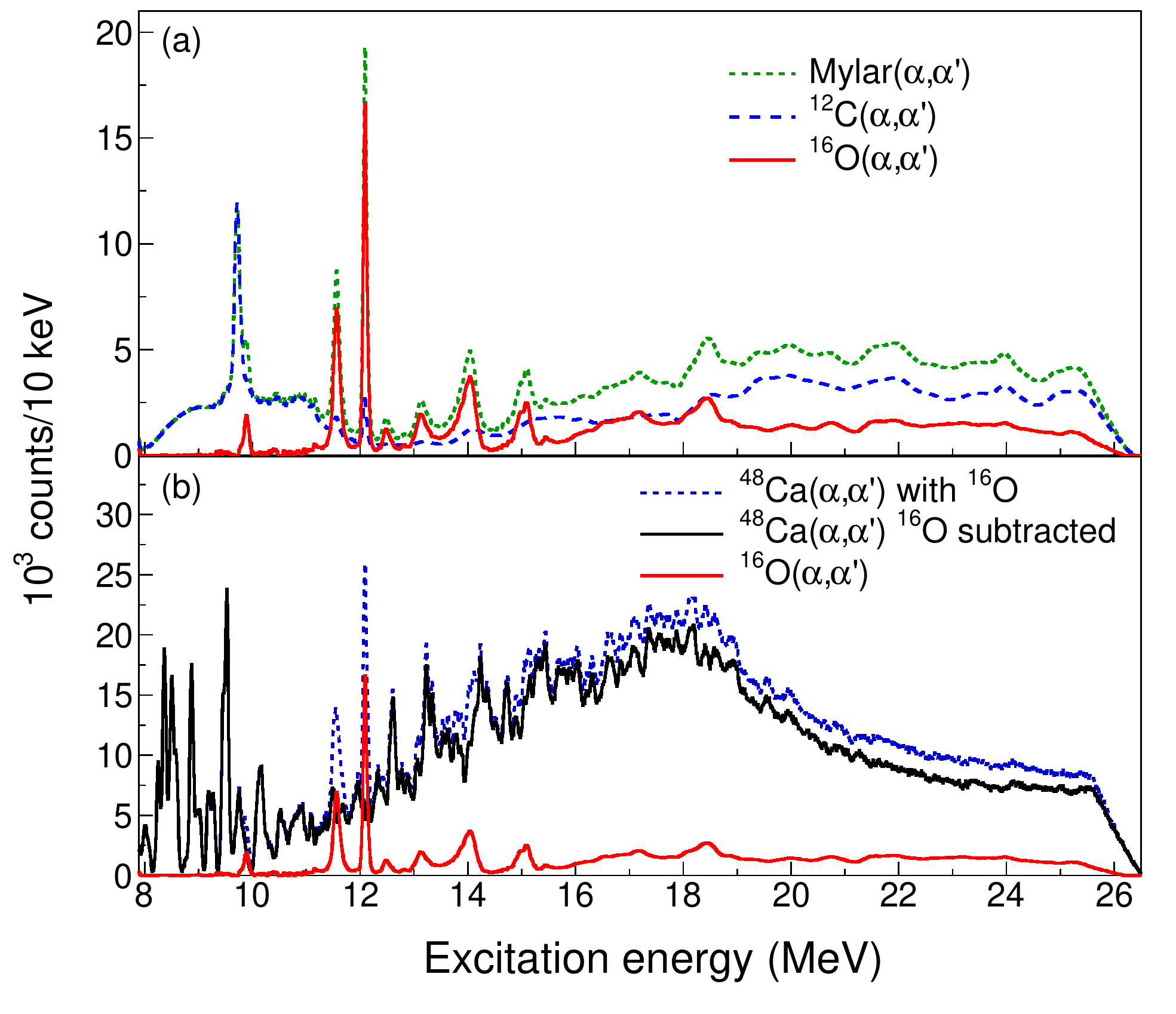}
\caption{
(a) Background-subtracted excitation-energy spectra for inelastic $\alpha$-particle 
scattering  at $\theta$\textsubscript{lab} = 0$^{\circ}$ 
off Mylar (short-dashed green line) and $^{12}$C (long-dashed blue line). 
The solid red line shows the difference between the Mylar and normalized 
$^{12}$C spectra, and represents
the excitation-energy spectrum for the $^{16}\textrm{O}$($\alpha,\alpha^\prime$) reaction. 
(b) Background subtracted excitation-energy spectrum for the oxygen-contaminated 
$^{48}$Ca target (dashed blue line), the normalized $^{16}\textrm{O}$  
spectrum (solid red line) and oxygen-free $^{48}$Ca spectrum (solid black line).
}
\label{fig:OxygenBackground} 
\end{figure}

Calcium is readily oxidized, and the presence of an $^{16}\textrm{O}$ contaminant on all the calcium targets  was identified by observing the prominent 11.520 MeV $J^{\pi}=2^+$ and 12.049 MeV $J^{\pi}=0^+$  states from $^{16}\textrm{O}$.
In order to extract accurate ISGMR strengths for the calcium isotopes the contribution of
this contaminant must be subtracted.
To this end, inelastic $\alpha$-particle scattering data from Mylar ($\textrm{C}_{10}\textrm{H}_8\textrm{O}_4$)  and $^{nat}\textrm{C}$ targets were acquired at $\theta$\textsubscript{lab} = 0$^{\circ}$ and 4$^{\circ}$.
An excitation-energy spectrum for the $^{16}\textrm{O}$($\alpha,\alpha^\prime$) reaction at each angle 
was then produced by subtracting the $^{12}$C data from the Mylar spectrum, normalized to the 9.641 MeV $J^{\pi}=3^-$ state and the broad excitation-energy region below it.
The contribution of the $^{16}\textrm{O}$ contaminant to the excitation-energy spectrum
of each calcium isotope was then removed by subtracting a normalized $^{16}\textrm{O}$ spectrum.
In the case of the 0$^{\circ}$ dataset the normalization was based on the strength of the $^{16}\textrm{O}$ 12.049 MeV $J^{\pi}=0^+$ peak.
For the 4$^{\circ}$ dataset the normalization was achieved using the strength of the 
$^{16}\textrm{O}$ 11.520 MeV $J^{\pi}=2^+$ peak.
Relevant sample spectra for the 0$^{\circ}$ case are shown in Fig.~\ref{fig:OxygenBackground}.

Well-known states in $^{24}$Mg were used for energy calibration of the focal plane.
An energy resolution of $\Delta E$ = 85 keV full width at half maximum (FWHM) was achieved, which allows for the 
observation of pronounced fine structure (a global phenomenon expected in giant resonances \cite{vnc19}) for the case of the ISGMR,
hitherto unseen in other inelastic $\alpha$-particle scattering measurements for the calcium isotope chain.

\section{\label{sec:results}RESULTS}

\subsection{\label{subsec:DoS}DoS technique}

The double-differential cross sections obtained in the 0$^{\circ}$ measurements 
can be ascribed predominantly to the monopole strength.
However, non-negligible contributions from other multipolarities are present and should be taken into account.
This can be achieved by employing the MDA technique \cite{Gup16b,Lui11},  which nominally requires data at numerous scattering angles.
Alternatively, in the absence of data at a large number of scattering angles, the difference-of-spectra (DoS) method can be applied. In this technique, utilized in the present study, the ISGMR strength is isolated from other multipoles contributing to the cross section measured in the 0$^{\circ}$ mode by subtracting the 
double-differential cross sections 
as observed for a small angular range representing the first minimum of the ISGMR angular distribution, from the 0$^{\circ}$ excitation-energy cross section distribution \cite{Bra87,You97,Gup16}.
This difference spectrum can then be considered representative of only the monopole strength
if one assumes that the angular distribution of the cross section for the sum of all multipolarities, but excluding the ISGMR, is flat and featureless for angles up to the first minimum of the ISGMR.

\begin{figure}
\centering
\includegraphics[width=1\columnwidth]{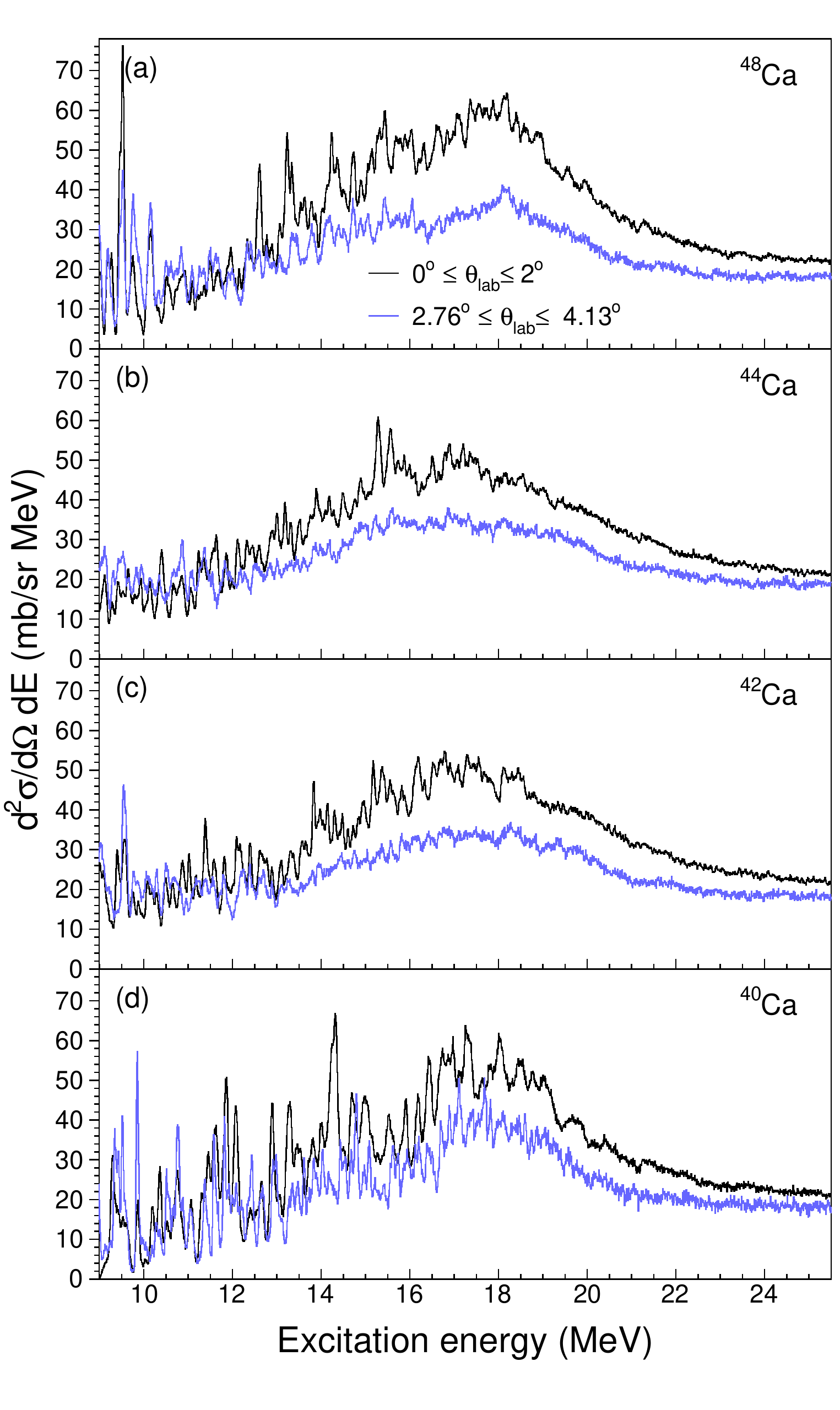}
\caption{
Measured double-differential cross sections for inelastic $\alpha$  scattering from $^{40,42,44,48}$Ca at $E_{\alpha}$ = 196 MeV.  
The black (top) line in panels (a) - (d) represents data for the angle range $0^{\circ}\leq\theta$\textsubscript{lab} $\leq 2^{\circ}$,
while the blue (bottom) correspond to an angular range
$2.76^{\circ}\leq\theta$\textsubscript{lab} $\leq 4.13^{\circ}$.
}
\label{fig:cs}
\end{figure}

\begin{figure}
\begin{center}
\includegraphics[width=1.\columnwidth]{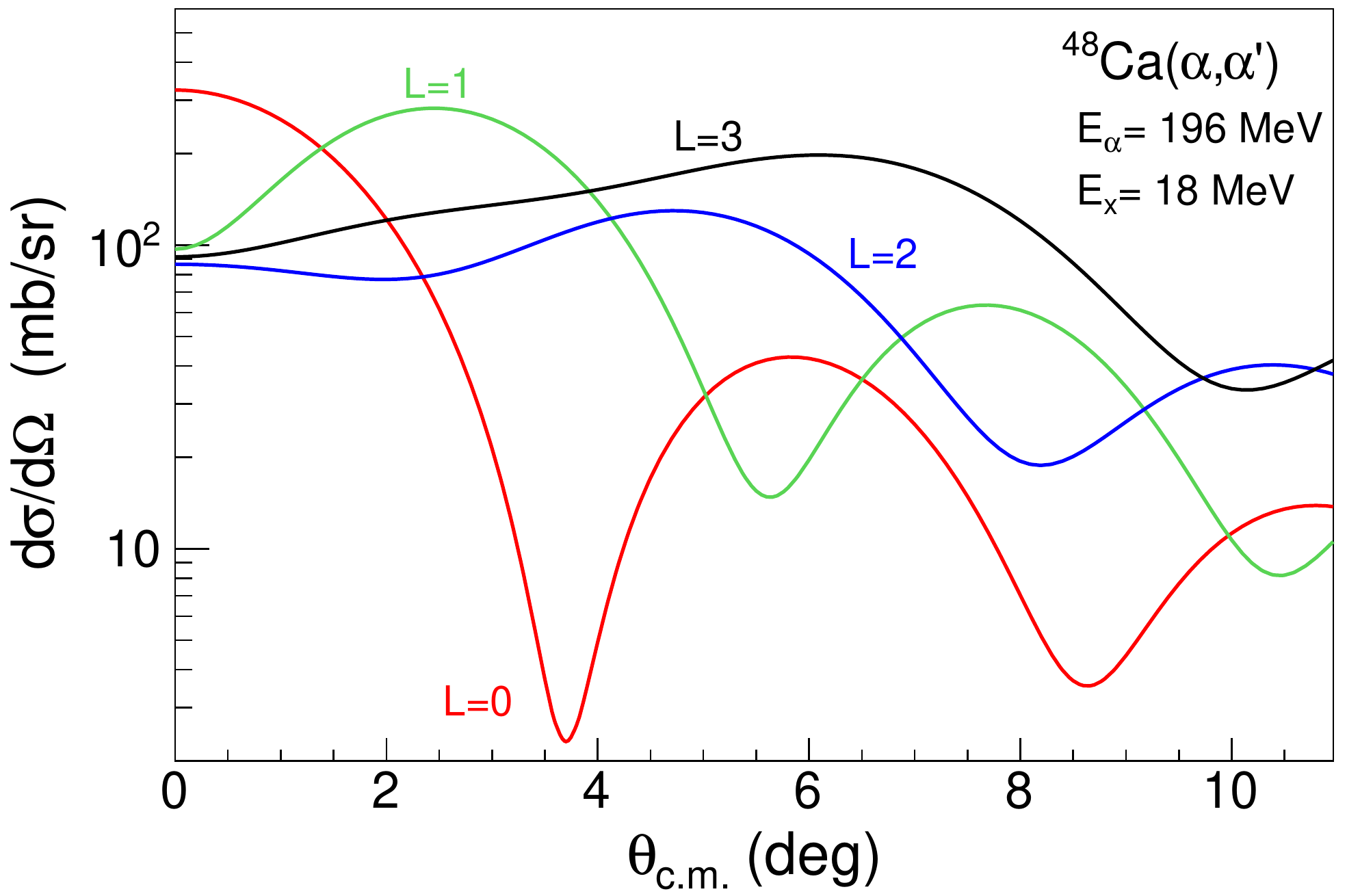}
\caption{DWBA calculated differential cross sections 
for inelastic $\alpha$ scattering on 
$^{48}$Ca at a beam energy of $E_{\alpha} = 196$ MeV 
for various isoscalar electric multipoles. 
The calculations were normalized to 100\% of the EWSR for each multipole 
at an excitation energy of 18 MeV in $^{48}$Ca, representing the peak of the cross section distribution.
}
\label{fig:dwba48Ca}
\end{center}
\end{figure}

\begin{figure}
\includegraphics[width=1.0\columnwidth]{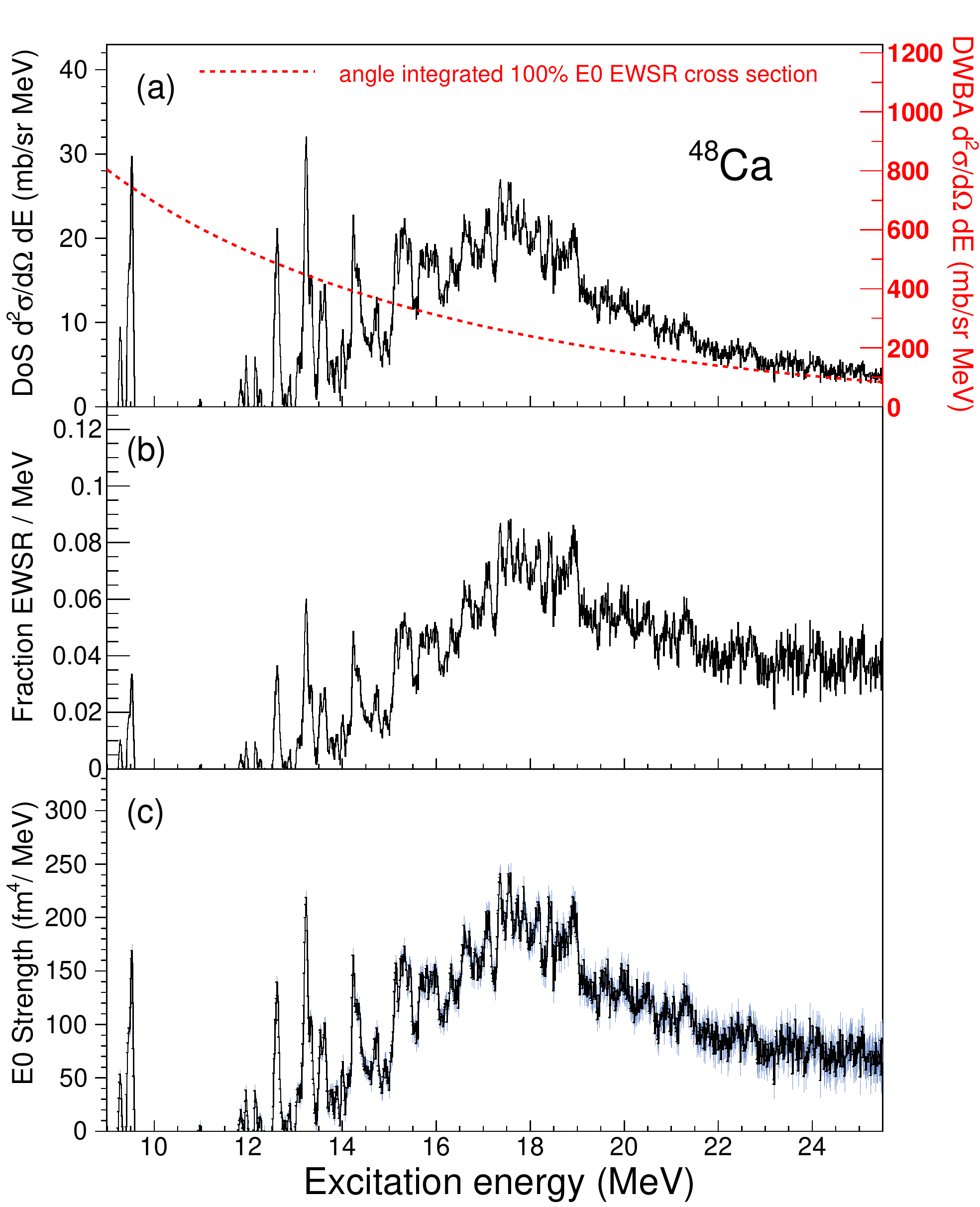}
\caption{
Outline of the conversion process from the $^{48}$Ca difference cross sections (a), 
to the fractional EWSR strength per MeV (b), and then to the $^{48}$Ca $E0$ strength distribution (c).
The red dashed line in panel (a) represents the 
$\theta$\textsubscript{lab} $\leq 2^{\circ}$
angle-integrated $L = 0$ DWBA cross section for 100\% of the $E0$ EWSR. 
Statistical errors are indicated in blue for the $E0$ strength distribution. }
\label{fig:DoSsimple}
\end{figure}

Double-differential cross sections for 0$^{\circ}$ ($0^{\circ}\leq\theta$\textsubscript{lab} $\leq 2^{\circ}$) 
and finite-angle ($2.76^{\circ}\leq\theta$\textsubscript{lab} $\leq 4.13^{\circ}$) measurements are shown in Fig.~\ref{fig:cs} for the four calcium isotopes under investigation.
The finite angle range spans the first minimum of the ISGMR cross section 
established through distorted-wave Born approximation (DWBA) calculations 
equivalent to $3^{\circ}\leq\theta$\textsubscript{c.m.} $\leq 4.5^{\circ}$, as shown in
Fig.\ref{fig:dwba48Ca}.
The DWBA calculations were performed with the nuclear reaction code PTOLEMY, 
utilizing the transition density and sum rule for the ISGMR breathing-mode oscillation as described
by Satchler and Khoa \cite{Sat97}. 
The optical model potentials 
were constructed using the hybrid single-folding approach, where the imaginary volume potential takes a Woods-Saxon form, while the real potential is obtained by folding the ground-state density with a density-dependent $\alpha$-nucleon interaction \cite{Sat97}.
Values for the optical model parameters are available from Ref.~\cite{Lui11} for the $^{48}$Ca nucleus at an incident beam energy of 240 MeV.  
Similar to Ref.~\cite{But17}, we used the same optical model parameters for the different calcium isotopes. 
While optical model parameters are also available for the $^{40}$Ca nucleus at 240 MeV \cite{You97} we opted to utilize the same parameter set for all the isotopes for the sake of consistency, since large differences are observed between the strength, radius, and diffuseness parameters of the $^{48}$Ca and $^{40}$Ca parameter sets. 
Nevertheless, test calculations were also performed with the parameters of Ref.~\cite{You97}, and the differences of resulting ISGMR strengths were found to be minimal.

\begin{figure*}
\includegraphics[width=1.0\textwidth]{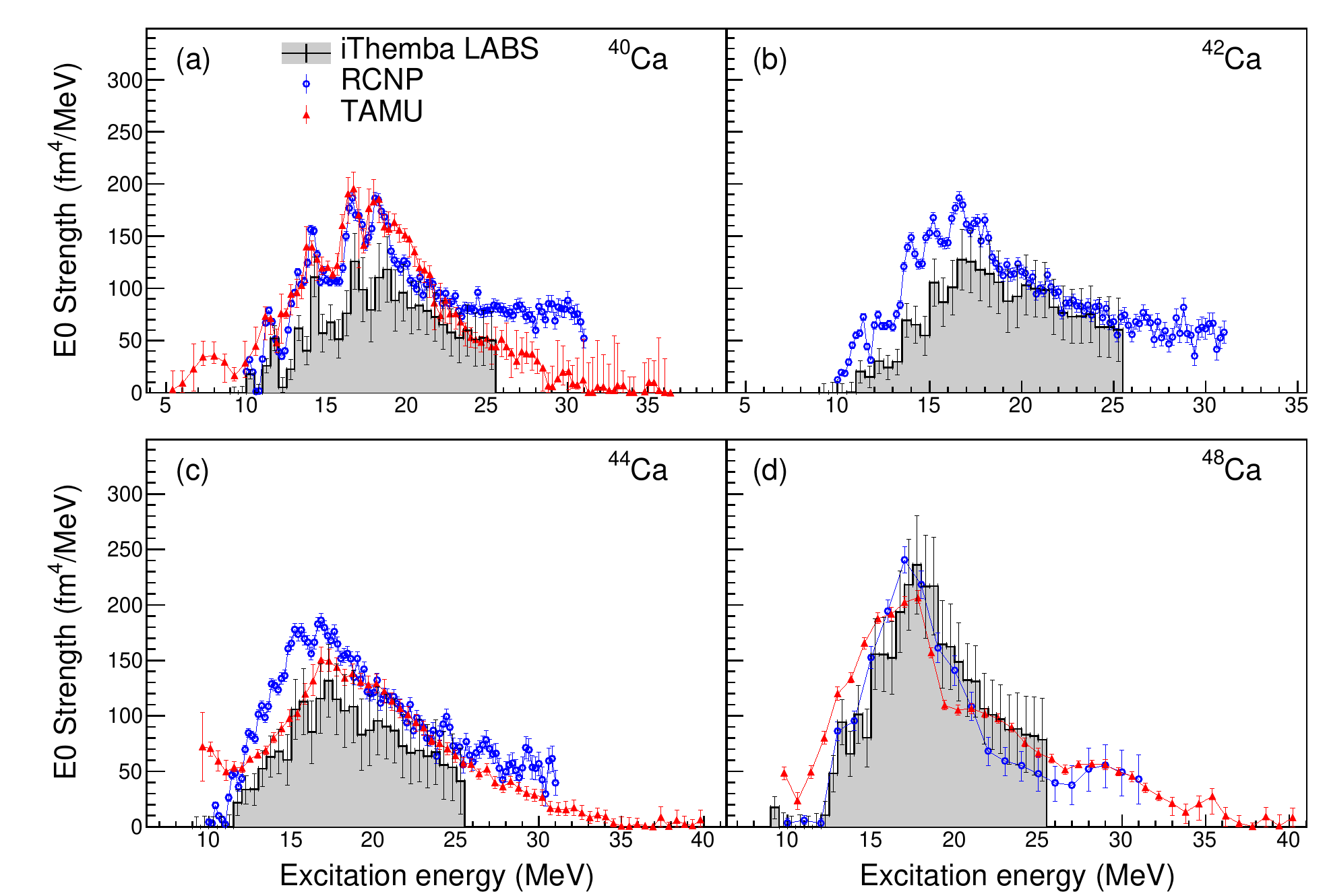}
\caption{ISGMR strength distributions for $^{40,42,44,48}$Ca from this study (gray filled histograms) utilizing the basic DoS technique, compared with results from RCNP \cite{HOWARD2020135185} (blue circles) and TAMU \cite{You01, Lui11, But17} (red triangles). 
The error bars of the present work represent both statistical and systematic uncertainties.
}
\label{fig:E0strengthTamuRCNPiTLsimple}
\end{figure*}
Extraction of the monopole strength from this work is limited to the excitation-energy range 9.5 - 25.5 MeV.
This represents the region where data for the full angular acceptance
are available for both the zero-degree and finite-angle double-differential cross section distributions.
The direct subtraction of these distributions for each isotope yields the difference spectra.
The evolution of the results from the difference spectrum 
to the final $E0$ strength distribution is illustrated in Fig.~\ref{fig:DoSsimple} for the case of $^{48}$Ca.
The energy-weighted sum rule (EWSR) fraction of the ISGMR strength 
as a function of excitation energy, represented by the symbol $a_0(E_x)$,
was determined by dividing the difference cross section (converted from the laboratory to the center-of-mass reference frame) by the $L = 0$ DWBA cross section distribution corresponding to 100\% of the ESWR averaged over the $0^{\circ}\leq\theta$\textsubscript{lab} $\leq 2^{\circ}$ angle range.
The final ISGMR strength distributions $S_0(E_x)$ were extracted using the expression
\begin{equation}
S_0(E_x) = \frac{2\hbar^2A\langle r^2\rangle}{mE_x}a_0(E_x)  \textrm{  ,}
\label{eq:EWSR}
\end{equation}
where $m$ and $A$ refer to the nucleon mass and mass number, respectively.
The symbol $\langle r^2\rangle$ refers to the moment of the ground state density, assuming
a Fermi mass distribution with half-density radius and diffuseness parameters 
taken from Ref.~\cite{Fri95}, as also used in previous studies \cite{HOWARD2020135185}.

The resulting isoscalar monopole strength distributions for $^{40,42,44,48}$Ca are shown in
Fig.~\ref{fig:E0strengthTamuRCNPiTLsimple}.
The error bars represent a combination of the statistical and 6.5\% systematic uncertainties associated with each of the double-differential cross section results.
Results from the studies at RCNP \cite{HOWARD2020135185} and TAMU \cite{You01, Lui11, But17} are shown for comparison. In the latter case the published fractional EWSR strengths were converted to $E0$ strength distributions in the manner explained above.
In the comparison it is clear that there are significant differences in the 
excitation-energy ranges covered. 
While the present results only cover the range 9.5 - 25.5 MeV, the studies performed at TAMU
presented $E0$ strengths for significantly larger excitation-energy ranges. For $^{48}$Ca  \cite{Lui11}, $^{44}$Ca \cite{But17} and $^{40}$Ca \cite{You01} these ranges are 9.8 - 40.2 MeV, 9.6 - 40 MeV and 5.4 - 36.4 MeV, respectively.
On the other hand, the measurements at RCNP  
were all done for an excitation-energy range 10 - 31 MeV \cite{HOWARD2020135185}.

Nevertheless, it is clear that for the three nuclei $^{40,42,44}$Ca one observes that the monopole strength distributions from the current study are weaker than those from RCNP and TAMU, especially at excitation energies around and below the peak of the distribution. The case of $^{48}$Ca, however, is somewhat different, and our results are in better agreement with previous datasets, although still weaker than the results from TAMU below the peak of the distribution.

\subsection{\label{subsec:DoSEx}DoS technique with excitation-energy-dependent correction}

\begin{figure}[t]
\includegraphics[width=1.\columnwidth]{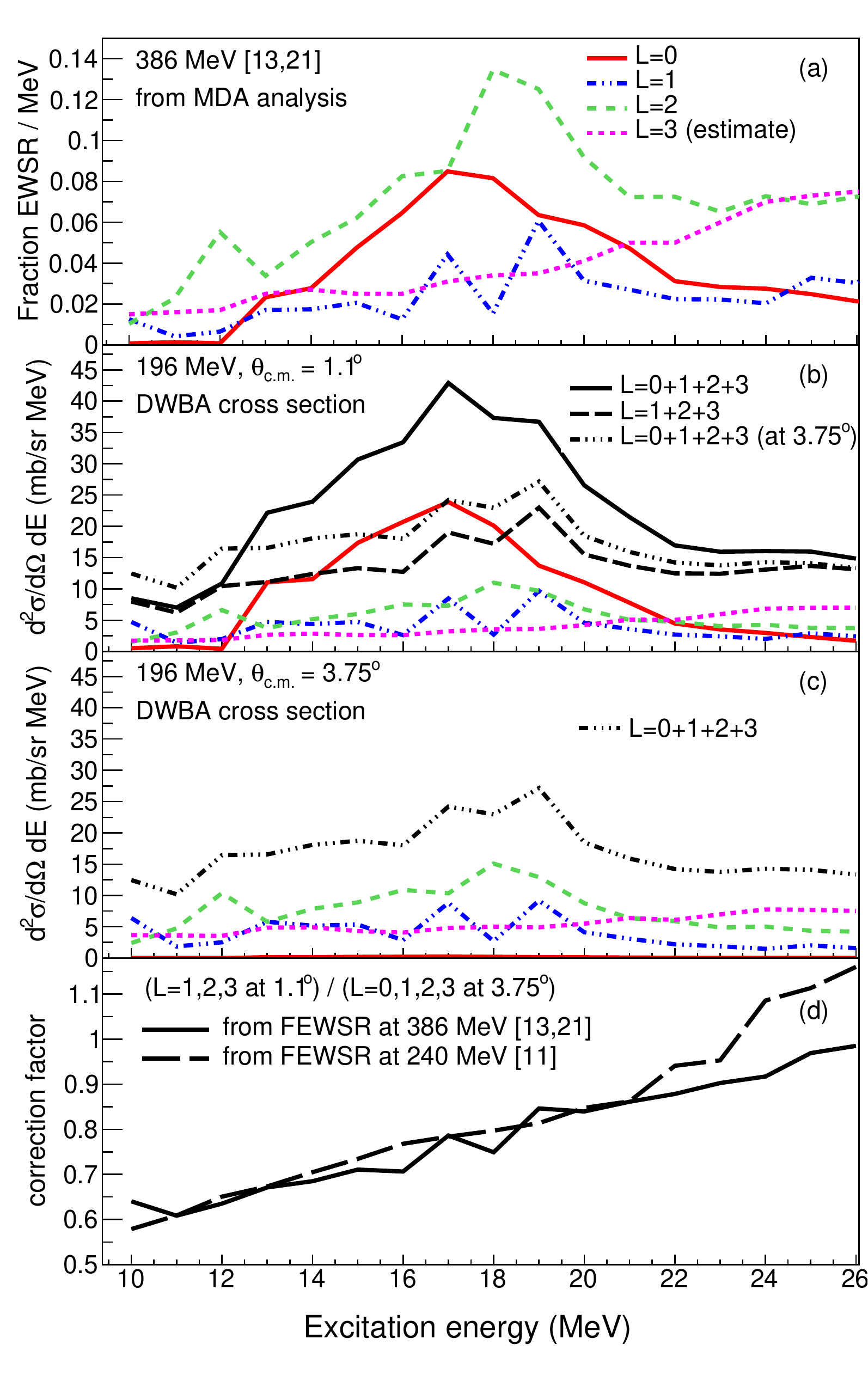}
\caption{
Outline of the procedure to establish an excitation-energy-dependent correction factor for the finite-angle cross sections, taking $^{48}$Ca as an example.
The fractional EWSR strengths for different multipolarities from RCNP  \cite{howard2020thesis,HOWARD2020135185} (a)
are used to determine DWBA cross sections at 196 MeV representative of the zero degree (b) and the
finite-angle measurements (c). The correction factors (solid line) shown in panel (d) are determined by the ratio of the $L=1,2,3$ results in panel (b) to the $L=0,1,2,3$ results in panel (c). The dashed line in panel (d) represents the correction factors when using the fractional EWSR results from TAMU \cite{Lui11}.
}
\label{fig:ratio48CaRCNP}
\end{figure}

\begin{figure}[ht]
\includegraphics[width=1.\columnwidth]{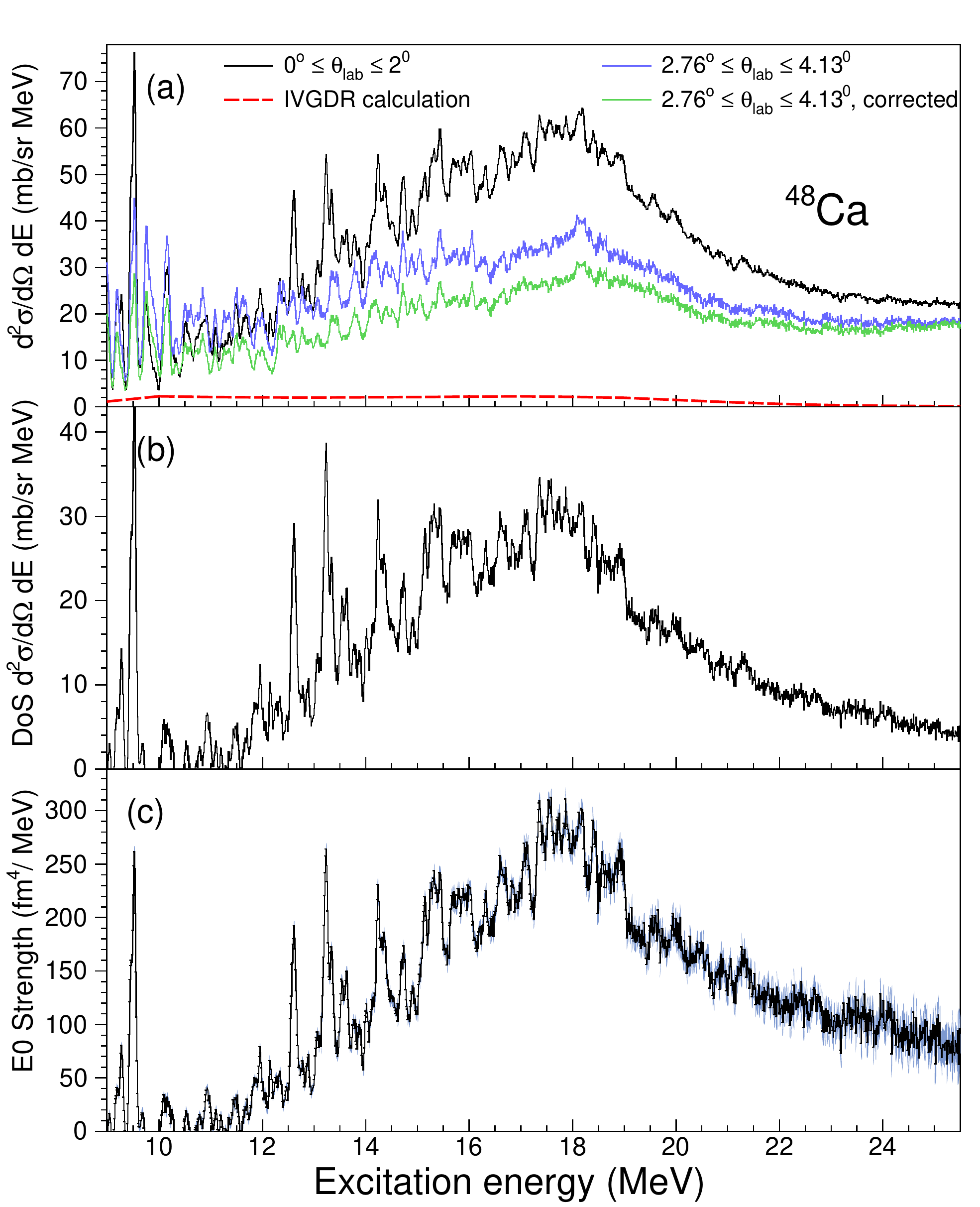}
\caption{(a) Application of the DoS method with excitation-energy-dependent correction factors for the case of $^{48}$Ca. The finite-angle cross section spectrum (dotted blue line) is modified with the RCNP-based excitation-energy-dependent correction factor from panel (d) in Fig.~\ref{fig:ratio48CaRCNP}, resulting in the solid green line. The weak IVGDR contribution (dashed red line) is subtracted from the zero-degree spectrum (solid black line), with the resulting DoS cross section shown in panel (b).
The $E0$ strength distribution in panel (c) can be compared with the basic DoS result in panel (c) of Fig.~\ref{fig:DoSsimple}. Statistical errors are indicated in blue for the $E0$ strength distribution. 
}
\label{fig:withGDR}
\end{figure}

Given the apparent lack of agreement it is necessary to reconsider the assumptions inherent to the method used in this study to calculate the $E0$ strength.
In particular, to what degree can the $L>0$ contributions in the 0$^{\circ}$ measurement 
be approximated by the cross section from the finite angle range centered around the first minimum of the $L=0$ component?
Note that the DoS technique is based merely on the general observation that while the $L=0$ strength is very forward peaked with a deep first minimum at small angles, all the other multipoles exhibit a much weaker angular dependence in the angle range around the first minimum. This much is clear from Fig.~\ref{fig:dwba48Ca}.
The technique was therefore originally conceived as a way to access monopole strength  
in the absence of other methods, 
but the degree to which it can be used for the detailed extraction of the monopole strength 
must be established.

To gauge the central premise of the DoS technique, specifically for the calcium isotope chain,
we use the strength distributions as determined in the MDA study of Howard {\it et al.}~\cite{HOWARD2020135185,howard2020thesis}. 
The $L=0,1,2$ strengths were converted to fractions of the EWSR for the various multipole strengths as a function of excitation energy, $a_L(E_x)$.
For the case of higher multipoles, we assumed that the cross section of the  $L=3$ component sufficiently resembles the sum of higher multipoles in the small angle range we are interested in, 
and estimated the fraction of the ESWR by scaling the $L=3$ DWBA cross section at 386 MeV to similar levels as the sum of higher multipole results in the angular distributions for various excitation energies published in Ref.~\cite{howard2020thesis}.
These  $L=0,1,2,3$ EWSR fractions were then applied to the result of DWBA calculations for 100\% EWSR as a function  of  excitation energy at an incident beam energy of 196 MeV. This procedure was first done for the angle $\theta\textsubscript{c.m.}$ = 1.1$^{\circ}$ to represent the average of the 0$^{\circ}$ measurement, and then at $\theta\textsubscript{c.m.}$ = 3.75$^{\circ}$ which represents the angle at which the monopole contribution is minimal.
This allows one to compare the estimated portion of the cross section 
 for the 0$^{\circ}$ measurement ($0^{\circ}\leq\theta$\textsubscript{lab} $\leq 2^{\circ}$) 
 consisting of all contributions except $L=0$, with an estimate for the sum of all multipoles from the finite-angle ($2.76^{\circ}\leq\theta$\textsubscript{lab} $\leq 4.13^{\circ}$) measurement. 
 The results for the case of $^{48}$Ca are shown in Fig.~\ref{fig:ratio48CaRCNP}. Panel (b) 
 demonstrates that the cross sections from the finite-angle measurement overestimate the sum of all $L>0$ multipoles in the 0$^{\circ}$ measurement. This overestimation decreases with increasing excitation energy.
 
 An excitation-energy-dependent correction factor can consequently be extracted and applied to the finite-angle cross sections to ensure the proper subtraction of the contributions of $L>0$ components to the zero-degree cross sections.
 As can be seen in panel (d) of Fig.~\ref{fig:ratio48CaRCNP} a very similar correction function was extracted from the fraction EWSR results obtained at TAMU \cite{Lui11}.
 Comparable agreement of correction functions deduced from the RCNP and TAMU datasets was also found for the cases of $^{40}$Ca and $^{44}$Ca.
 While it is therefore true that this correction procedure makes the present results model dependent, it does not bias the results in order to agree with any particular previously observed result from either RCNP or TAMU.
 
\begin{figure*}
\includegraphics[width=1.0\textwidth]{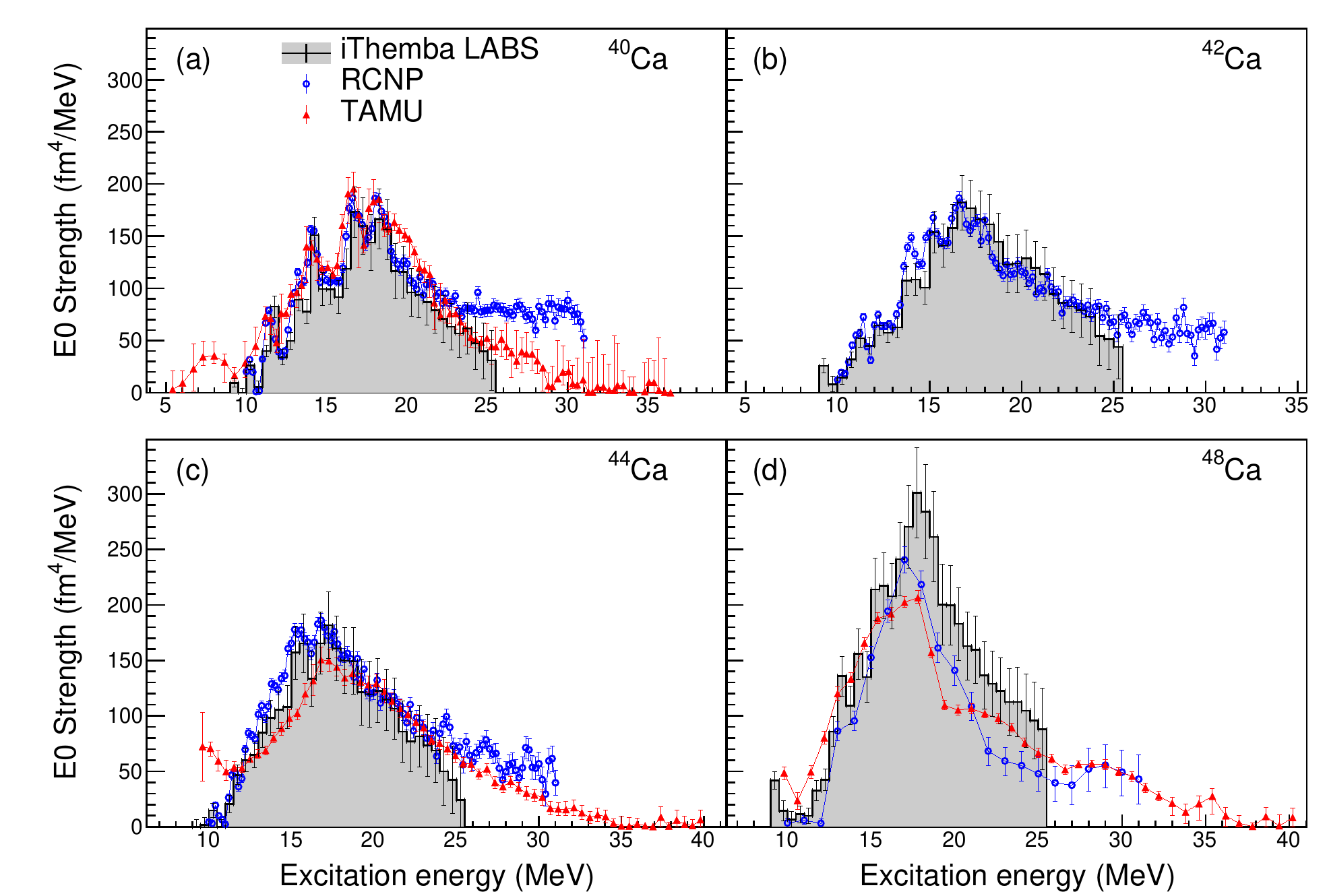}
\caption{Same as Fig.\ref{fig:E0strengthTamuRCNPiTLsimple} but using excitation-energy-dependent correction factors as described in the text.
}
\label{fig:E0strengthTamuRCNPiTL}
\end{figure*}

 In panel (a) of Fig.~\ref{fig:withGDR} the effect of the correction factor on the finite-angle cross sections is illustrated. In addition, a possible contribution from the isovector giant dipole resonance (IVGDR), estimated using the Goldhaber-Teller model and photonuclear cross section data \cite{SATCHLER1987215, PLUJKO20181}, is considered.
 Such a contribution would violate the basic assumptions of the DoS and must be removed prior to the subtraction, since the angular distribution is strongly forward peaked.
 As seen here, and also known from from previous work \cite{Shlomo1987}, the contribution from the IVGDR is rather small in lighter nuclei, although non-zero.
 The difference spectrum shown in panel (b) of Fig.~\ref{fig:withGDR} is the result of the subtraction of the corrected finite-angle double-differential cross section as well as the IVGDR contribution from the zero-degree double-differential cross section.   
 Following the same procedure discussed earlier to convert to the $E0$ strength from the difference spectrum, we now extract significantly increased values for $S_0(E_x)$, which are presented in Fig.~\ref{fig:E0strengthTamuRCNPiTL}.
 
\section{\label{sec:Discussion}Discussion}

It is clear 
for the case of calcium isotopes that
the assumption of varying contributions from different $L>0$  multipoles 
collectively resulting in a completely flat angular distribution across a wide 
excitation-energy region is not accurate.
Once proper allowance is made for the real angular character of the $L>0$  multipoles
we find that the DoS technique yields monopole strength distributions which are in good agreement with previous results from Howard {\it et al.} \cite{HOWARD2020135185} at 386 MeV for $^{40,42,44}$Ca.  
In the comparison with 
the TAMU results 
we observe fair agreement in the case of 
$^{40}$Ca, but see marked structural differences for $^{44}$Ca. 
However, our result for $^{48}$Ca does not agree well with either previous result.  
While it is seen, similar to $^{40,44}$Ca, that at low excitation energies (below $\sim$12 MeV) the strengths agree with the RCNP dataset in being lower than the TAMU results, we find higher values than seen before for the strength distribution at its peak and beyond,  and similar to the TAMU results the peak of the distribution is also found at a higher excitation energy than the RNCP result.

\begin{figure}[t]
\includegraphics[width=1\columnwidth, angle=0]{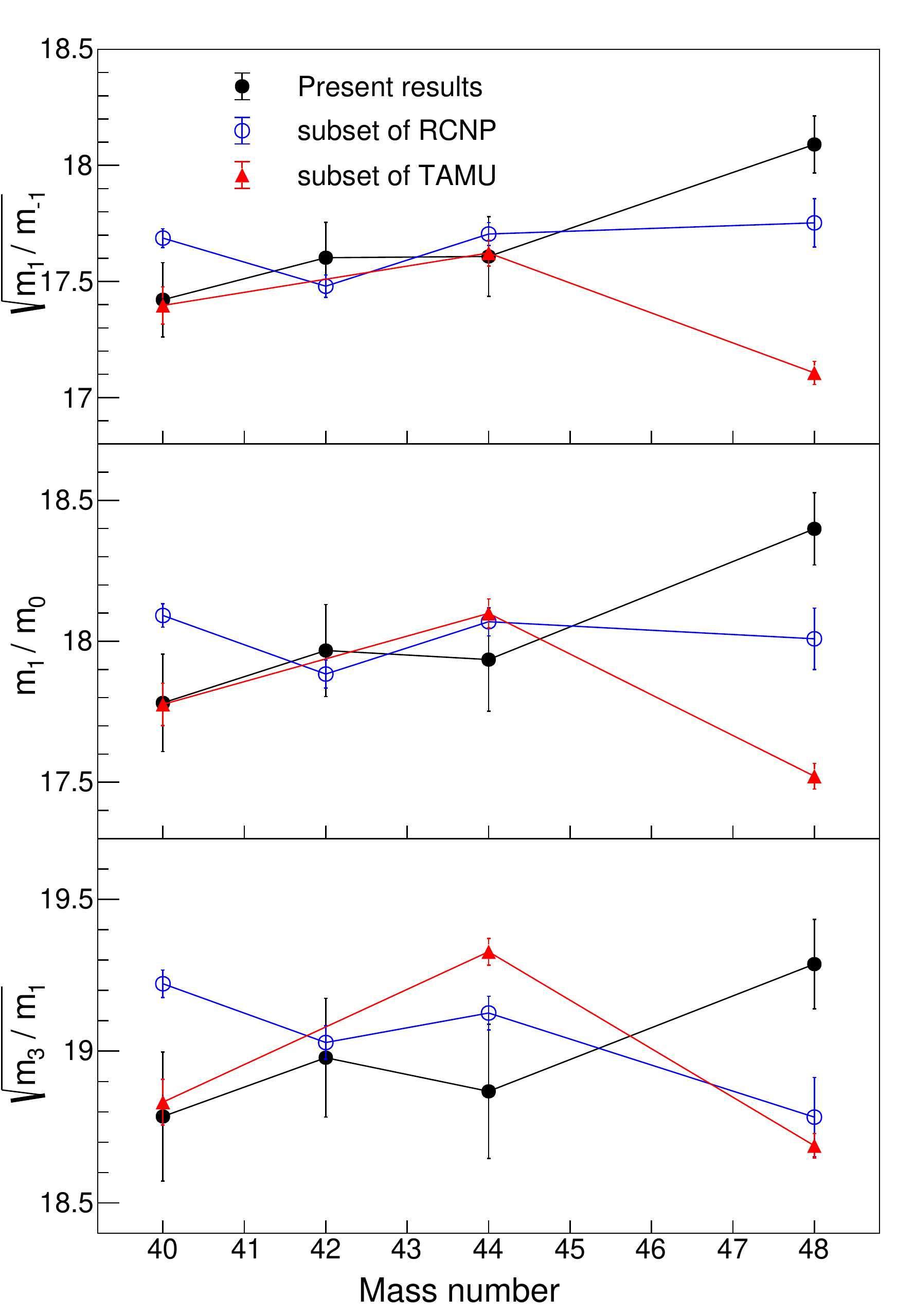}
\caption{Moment ratios of the $E0$ strength distributions in $^{40,42,44,48}$Ca for the excitation-energy range 10 - 25.5 MeV from the present work (black circles), compared to values extracted for the TAMU \cite{You01, Lui11, But17} (red triangles) and RCNP \cite{HOWARD2020135185} (blue empty circles) datasets over the same excitation-energy range. }
\label{fig:momentratios}
\end{figure}

While the results from this work do  not  provide  access  to a sufficiently extended excitation-energy range to allow for direct comparisons with published values of $K_A$, it can help us gain insight into the origin of the different systematic trends observed.
For the data from iThemba LABS, RCNP and TAMU the moments of the monopole strength distributions
\begin{equation}
m_k = \int S_0(E_x) E^k_x dE_x \textrm{  ,}
\label{eq:moment}
\end{equation}
were obtained for an excitation-energy range common to all three sets of results, namely $10 - 25.5$ MeV.
The resulting moment ratios $\sqrt{m_1/m_{-1}}$, $m_1/m_0$ and $\sqrt{m_3/m_{1}}$, customarily used to characterize the energy of the ISGMR, are summarized in Table \ref{table:momentratios} and shown in Fig.~\ref{fig:momentratios}.
Error bars were calculated using Monte Carlo sampling of the $E0$ strength distributions and their associated error bars in the calculation of the moments and moment ratios, and represent a 68\% confidence interval.
Fair agreement between the different datasets, and no mass dependent systematic trend, is observed for the moment ratios $m_1/m_0$ and $\sqrt{m_1/m_{-1}}$ for the nuclei $^{40,42,44}$Ca. However, in the case of $^{48}$Ca the prominent shape differences in the strength distributions are reflected in differences between the moment ratios from the three experiments. As can be intuited from panel (d) of Fig.~\ref{fig:E0strengthTamuRCNPiTL}, the moment ratios for $^{48}$Ca will be largest for the iThemba LABS dataset and smallest for the TAMU dataset.
Furthermore, for the iThemba LABS dataset these moment ratios for $^{48}$Ca are consistently higher than those for the other three isotopes, while the opposite is true in the case of the TAMU dataset.  For the RCNP dataset the values of the $^{48}$Ca moment ratios agree within 
uncertainty with results of the other calcium isotopes.

For the case of the scaling energy $\sqrt{m_3/m_1}$ the higher weight assigned to the strength distribution at higher excitation energies results in a weak downward sloping trend for the RCNP dataset as mass is increased. The scaling energy calculated for the TAMU dataset follows the RCNP results closely for $^{44}$Ca and $^{48}$Ca, while being much lower in the case of $^{40}$Ca. The iThemba LABS scaling energy results are, within error, rather flat.

\begin{table}[!ht]
\caption{
Moment ratios  of the $E0$ strength distributions in $^{40,42,44,48}$Ca calculated over the excitation-energy range 10 - 25.5 MeV. 
\\ \vspace{2mm}}
\label{table:momentratios}
\centering
\begin{ruledtabular}
\begin{tabular}{cccc}
Nucleus   &\quad  This work &\quad  RCNP \cite{HOWARD2020135185}             &\quad  TAMU \cite{You01,Lui11,But17} \\ 
\hline 
          &\quad            &\quad $m_1/m_0$ (MeV)    &\quad  \\
\hline 
$^{40}$Ca &\quad 17.78 $\pm$ 0.17 &\quad 18.09 $\pm$ 0.05 &\quad 17.78 $\pm$ 0.10 \\ 
$^{42}$Ca &\quad 17.97 $\pm$ 0.16 &\quad 17.88 $\pm$ 0.05 &\quad - \\ 
$^{44}$Ca &\quad 17.94 $\pm$ 0.18 &\quad 18.07 $\pm$ 0.05 &\quad 18.10 $\pm$ 0.10 \\ 
$^{48}$Ca &\quad 18.40 $\pm$ 0.13 &\quad 18.01 $\pm$ 0.10  &\quad 17.52 $\pm$ 0.10 \\ 
\hline 
          &\quad            &\quad $\sqrt{m_1/m_{-1}}$ (MeV)    &\quad  \\
\hline 
$^{40}$Ca &\quad 17.42 $\pm$ 0.16 &\quad 17.69 $\pm$ 0.05 &\quad 17.40 $\pm$ 0.10 \\ 
$^{42}$Ca &\quad 17.60 $\pm$ 0.15 &\quad 17.48 $\pm$ 0.05 &\quad - \\
$^{44}$Ca &\quad 17.61 $\pm$ 0.17 &\quad 17.70 $\pm$ 0.05 &\quad 17.62 $\pm$ 0.10 \\ 
$^{48}$Ca &\quad 18.09 $\pm$ 0.12 &\quad 17.75 $\pm$ 0.10  &\quad 17.11 $\pm$ 0.10 \\ 
\hline 
          &\quad            &\quad $\sqrt{m_3/m_1}$ (MeV)    &\quad  \\
\hline 
$^{40}$Ca &\quad 18.79 $\pm$ 0.20 &\quad 19.22 $\pm$ 0.05  & \quad 18.83 $\pm$ 0.10 \\
$^{42}$Ca &\quad 18.98 $\pm$ 0.20 &\quad 19.03 $\pm$ 0.05  & \quad - \\
$^{44}$Ca &\quad 18.87 $\pm$ 0.22 &\quad 19.13 $\pm$ 0.05  & \quad 19.33 $\pm$ 0.05 \\ 
$^{48}$Ca &\quad 19.29 $\pm$ 0.15 &\quad 18.78 $\pm$ 0.13 &\quad 18.69 $\pm$ 0.05 \\ 
\end{tabular}
\end{ruledtabular}
\end{table} 

\begin{figure}[t]
\includegraphics[width=1\columnwidth, angle=0]{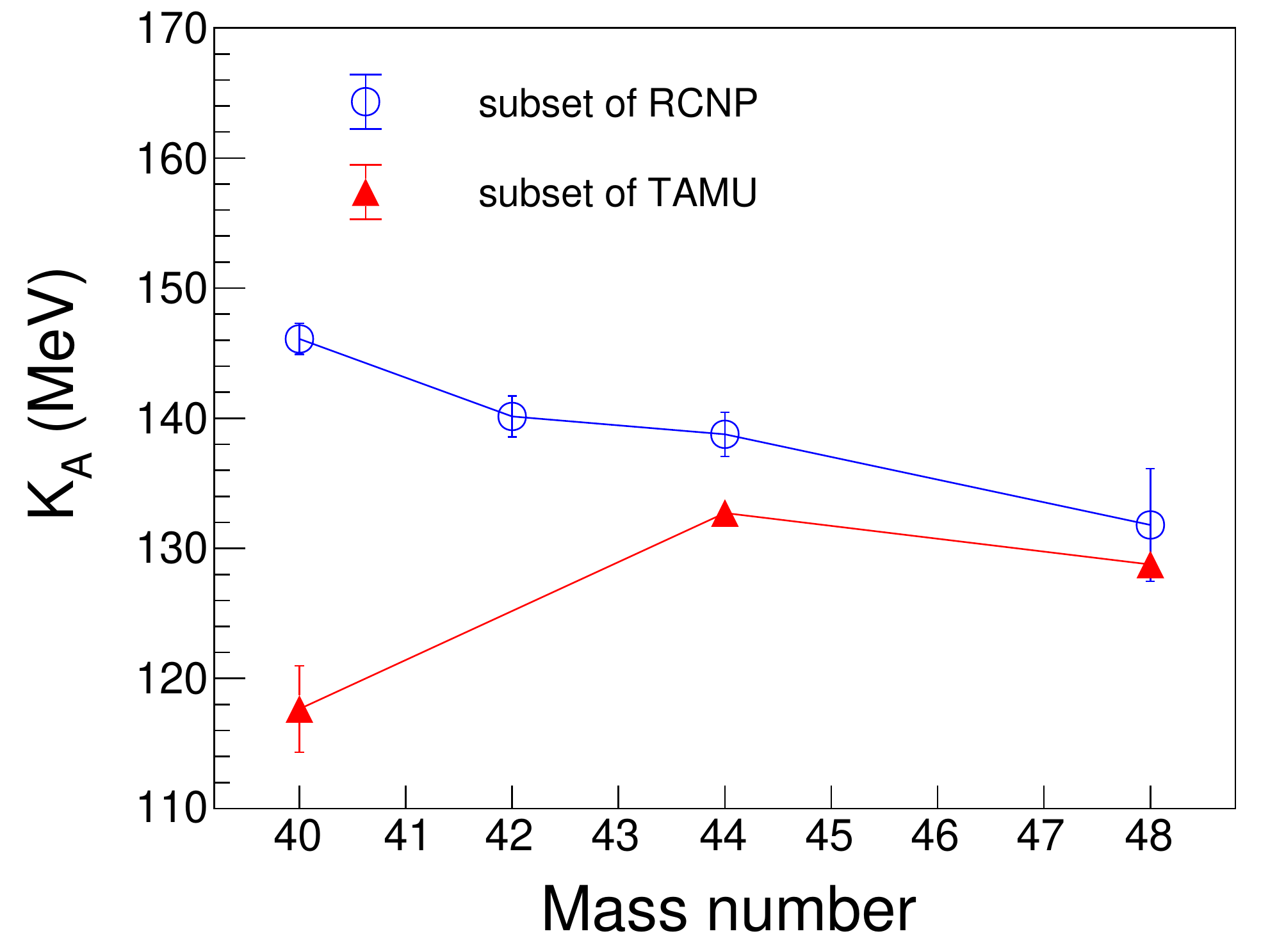}
\caption{Experimental results for the incompressibility of the nucleus 
$K_A$ 
for $^{40,42,44,48}$Ca extracted from the TAMU \cite{You01, Lui11, But17} (red triangles) and RCNP \cite{HOWARD2020135185} (blue circles) datasets over the excitation-energy range  10 - 31 MeV accessible at RCNP.}
\label{fig:KARCNPrange}
\end{figure}

Values of 
$K_A$ 
calculated using the scaling energy 
as representative of $E\textsubscript{ISGMR}$ such that
\begin{equation}
E\textsubscript{ISGMR} = \hbar \sqrt{ \frac{K_A}{ m \langle r^2 \rangle    } }  \textrm{  ,}
\label{eq:KA}
\end{equation}
where $m$ and $\langle r^2\rangle$ are defined in Eq.~(\ref{eq:EWSR}), would reveal the same weak trends as seen for the scaling energy.
However, when considering the value of $K_A$ 
for an excitation-energy range common to both TAMU and RCNP datasets, i.e.~10 - 31 MeV 
as seen in Fig. \ref{fig:KARCNPrange},
a downward trend is seen for the RCNP results, which is mirrored by the results from TAMU for  $^{44}$Ca and $^{48}$Ca.
The only conspicuous disagreement is found for the case of the $^{40}$Ca nucleus.
This is in noticeable contrast to the comparison when utilizing the full excitation-energy
region up to 40 MeV accessed in the TAMU experiment, 
as is shown in Ref.~\cite{HOWARD2020135185}, where the value of $K_A$ for $^{48}$Ca is dramatically increased, resulting in an overall upward trend 
with mass number.

The evolution of mass-dependent trends for the nuclear incompressibility 
extracted from the scaling energy as 
different excitation-energy regions are utilized, highlights the importance of the determination of the monopole strength at high excitation energies. This is clearly evident from the change in 
$K_A$ 
for  $^{40}$Ca in the RCNP and TAMU datasets when expanding the excitation-energy range from 10 - 25 MeV to 10 - 31 MeV. 
The differences 
of the 
$E0$ strengths in the region 25 - 31 MeV are responsible for the large relative changes  
in $K_A$  between the RCNP and TAMU results.
Furthermore, when the full excitation-energy range for the $E0$ strengths from TAMU is utilized, the non-zero values in the case of $^{48}$Ca at very high excitation energies up to 40 MeV (where no data are available for the RCNP dataset) are responsible for pulling up this particular value of $K_A$ to beyond that of the $^{44}$Ca nucleus, ultimately resulting in the contested increasing trend as a function of mass.
If we thus confine ourselves to the comparison of $K_A$ from previous studies for similar excitation-energy ranges, it would therefore seem that the source of previous opposing conclusions regarding the trend of $K_A$ for calcium isotopes does not necessarily stem from two different background subtraction methodologies affecting all nuclei, but specifically from the details of the background subtraction associated with one particular nucleus, namely $^{40}$Ca.

However, when comparing values for the nuclear incompressibility obtained from different experimental studies 
one should be cognisant of the fact that the value of $K_A$ is a single number
obtained from the ratio of moments of the $E0$ strength distribution, which displays quite a variation in structural character between different studies.  For example, 
for the case of $^{44}$Ca there exists an agreement in calculated values for $K_A$ despite clear structural differences. The agreement follows since the strength in the region 10 - 12 MeV where the TAMU dataset dominates is balanced out by the dominance of the iThemba LABS and RCNP $E0$ strengths in the excitation-energy range 12 - 17 MeV. 
In general the TAMU results exhibit an increase in $S_0(E_x)$ at lower excitation
energies that is mostly absent, or much decreased, in the results from this study and that of Howard {\it et al.} \cite{HOWARD2020135185},
while the bump between 5 and 10 MeV for the case of $^{40}$Ca claimed in Ref.~\cite{You01} cannot be 
independently confirmed here 
due to the lack of data in this energy region. 

These structural differences highlight the unsolved problems related to background subtraction in all existing datasets.
On the one hand the background subtraction methodology used in the TAMU experiments, which makes assumptions about both the instrumental background as well as physical continuum, is questioned in recent studies
\cite{Gup16,HOWARD2020135185}. 
On the other hand, while this study and the results from RCNP employs experimental methods that eliminates the experimental background, possible contributions due to the physical continuum are ignored. 
Consequently some portion of the extracted $E0$ strength, especially towards higher excitation energies, could be due to contributions to the continuum from three-body channels, such as knockout reactions and quasi-free processes \cite{Gar18}.
More work is clearly required to properly understand and subtract all background contributions.

\section{\label{sec:Conclusion}Conclusions}

In an effort to understand conflicting reports 
of the systematic trend of the ISGMR strength
in stable even-mass calcium isotopes, we present a third dataset acquired by inelastic $\alpha$ scattering
at an incident beam energy of 196 MeV.  
The main 
difference between the two previously published datasets, also acquired by means of inelastic $\alpha$ scattering at incident beam energies 240 and 386 MeV, pertains to the difference in approach to background subtraction.
The new measurement reported here employs a similar strategy to that followed
at RCNP, where the instrumental background can be measured and subtracted, as opposed to
the approximation applied in the case of TAMU.

While the present results  do  not  provide  access  to  sufficiently  
high excitation energies to allow for direct comparisons of the value of 
$K_A$ 
with 
previous studies, 
we find for all datasets that the moment ratios of the ISGMR strength confined to the excitation-energy range that defines the main resonance region, displays only a weak systematic sensitivity to the increase of mass. 
Different trends previously observed for the nuclear incompressibility are caused by contributions to the $E0$ strength outside of the main resonance region, in particular at high excitation energies.
While procedures exist to identify and subtract instrumental background, more work is required to characterize and subtract continuum background contributions at high excitation energies.

\section*{Acknowledgements}


The authors thank the accelerator staff at iThemba LABS for the provision of high-quality dispersion-matched $\alpha$-particle beams. 
%
S.O.~thanks Kevin Howard for providing the numerical results of the RCNP experiment. 
%
This work was supported by the National Research Foundation of South Africa and by the Deutsche Forschungsgemeinschaft under contract SFB 1245 (project ID 279384907).
P.J.~and R.N.~acknowledges financial support from the National Research Foundation of South Africa through Grant Nrs.~90741 and 85509.
P.A.~acknowledges support from the Claude Leon Foundation in the form of a postdoctoral fellowship.
%


\end{document}